\documentclass[twocolumn,10pt]{IEEEtran}
\usepackage{bbm,amssymb,amsmath,amsfonts,latexsym}
\usepackage{graphicx,color,subfigure}
\usepackage{tikz}
\usepackage{algorithm,algorithmic}
\usepackage{graphics,url}
\usepackage{multirow}
\usepackage{hyperref}
\definecolor{subtler}{rgb}{1,0,0.1}  % subtle shading
\newcommand{\be}{\begin{equation}}
\newcommand{\ee}{\end{equation}}
\newcommand{\ben}{\begin{equation*}}
\newcommand{\een}{\end{equation*}}
\newcommand{\ba}{\begin{eqnarray}}
\newcommand{\ea}{\end{eqnarray}}

\newcommand{\id}[1]{\mathbf{I}_{#1}}

\newcommand{\Real}{\mathbb R}

\newcommand{\I}{\boldsymbol{\mathcal{I}}_{\vth}}
\newcommand{\tr}{\mathbf{tr}}

\newcommand{\T}{\mathrm{T}}
\newcommand{\E}{\mathbb E}
\newcommand{\bEu}[1]{\mathbf{e}_{#1}}

\newcommand{\diag}{\mathbf{diag}}
\newcommand{\rank}{\mathrm{rank}}

\newcommand{\veth}{\boldsymbol{\hat \theta}}

\newcommand{\bxi}[1]{\mathbf{x}^{(#1)}}
\newcommand{\bXi}[1]{\mathbf{X}^{(#1)}}
\newcommand{\iomega}[1]{\omega^{(#1)}}
\newcommand{\bdi}[1]{\mathbf{d}^{(#1)}}
\def\nf{N_{\!f}} %   number of flows
\def\onen{\mathbf{1}_n}
\def\range{\text{range}}

\def\vth{\boldsymbol\theta}

\def\bmu{\boldsymbol{\mu}}
\def\bJ{\mathbf{J}_{\vth}}
\def\bc{\mathbf{c}_{\vth}}
\def\bW{\mathbf{W}_{\vth}}
\def\bG{\mathbf{G}_{\vth}}
\def\bA{\mathbf{A}}

\def\bH{\mathbf{H}_{\vth}}

\def\bP{\mathbf{P}}
\def\bW{\mathbf{W}_{\vth}}
\def\bU{\mathbf{U}_{\vth}}

\def\bX{\mathbf{X}}
\def\bY{\mathbf{Y}}
\def\bZ{\mathbf{Z}}
\def\bB{\mathbf{B}}
\def\bT{\mathbf{T}_{\vth}}
\def\bTheta{\boldsymbol{\Theta}}

\def\bx{\mathbf{x}}
\def\by{\mathbf{y}}
\def\bb{\mathbf{b}}

\def\iY{\boldsymbol{Y}}

\def\Spd{\mathbb{S}^{n}_{++}}
\def\Spsd{\mathbb{S}^{n}_{+}}

\graphicspath{{.}
{Figures/}
}

\def\twoup{80mm}

\begin{document}

\title{Computing Constrained Cram\'er-Rao Bounds}

\author{Paul~Tune,~\IEEEmembership{Member,~IEEE}
\thanks{The author is with the School of Mathematical Sciences, The University of Adelaide, Australia
(Email: paul.tune@adelaide.edu.au). Technical report version, TR01-2012.}}

\maketitle

%%%%%%%%%%%%%%%%%%%%%%%%%%%%%%%%%%%%%%%%%%%%%%%%%%%%%%%%%%%%%%%%
%%%%%%%%%%%%%%%%%%%%%%%%%%%%%%%%%%%%%%%%%%%%%%%%%%%%%%%%%%%%%%%%
\begin{abstract}
We revisit the problem of computing submatrices of the Cram\'er-Rao bound (CRB), which lower bounds
the variance of any unbiased estimator of a vector parameter $\vth$. We explore iterative methods that avoid 
direct inversion of the Fisher information matrix, which can be computationally expensive when the 
dimension of $\vth$ is large. The computation of the bound is related to the quadratic matrix program, where 
there are highly efficient methods for solving it. We present several methods, and show that algorithms in
prior work are special instances of existing optimization algorithms. Some of these methods converge
to the bound monotonically, but in particular, algorithms converging non-monotonically are much
faster. We then extend the work to encompass the computation of the CRB when the Fisher information
matrix is singular and when the parameter $\vth$ is subject to constraints. As an application, we
consider the design of a data streaming algorithm for network measurement.
\end{abstract}


%%%%%%%%%%%%%%%%%%%%%%%%%%%%%%%%%%%%%%%%%%%%%%%%%%%%%%%%%%%%%%%%
%%%%%%%%%%%%%%%%%%%%%%%%%%%%%%%%%%%%%%%%%%%%%%%%%%%%%%%%%%%%%%%%
\section{Introduction}
\label{sec:intro}

The Cram\'er-Rao bound (CRB) \cite{Kay93Estimation} is important in quantifying the best achievable
covariance bound on unbiased parameter estimation of $n$ parameters $\vth$. Under mild regularity conditions,
the CRB is asymptotically achievable by the maximum likelihood estimator. 
The computation of the CRB is motivated 
by its importance in various engineering disciplines: medical imaging \cite{Hero96BVtradeoff},
blind system identification \cite{BlindEst98}, and many others. 

A related quantity is the Fisher information matrix (FIM), whose inverse is the CRB. Unfortunately,
direct inversion techniques are known for their high complexity in space ($O(n^2)$ bytes of 
storage) and time ($O(n^3)$ floating point operations or flops). Often, one is just interested in
a portion of the covariance matrix. In medical imaging applications, for example, only a small
region is of importance, which is related to the location of a tumor or lesion. In this instance,
computing the full inverse of the FIM becomes especially tedious and
intractable when the number of parameters is large. In some applications, the FIM itself is singular, 
and the resulting Moore-Penrose pseudoinverse computation is even more
computationally demanding. Avoiding the additional overhead incurred from direct inversion or other
forms of matrix decompositions (Cholesky, QR, LU decompositions, for example) becomes a strong 
motivation. 

Prior work \cite{Hero94CRBComp,Hero96CRBRecursive} proves the tremendous savings in memory and
computation by presenting several recursive algorithms computing only submatrices of the CRB.
Hero and Fessler \cite{Hero94CRBComp} developed algorithms based on matrix 
splitting techniques, and statistical insight from the Expectation-Maximization (EM) algorithm. Only 
$O(n^2)$ flops are required per iteration, which is advantageous if convergence is rapid, and 
the algorithms produce successive approximations that converge monotonically to the CRB. Exponential 
convergence was reported, resulting in computational savings, with the asymptotic rate of convergence 
governed by the relationship between the FIM of the observation space and the complete 
data space. This seminal work was further extended in \cite{Hero96CRBRecursive}, where better 
choices of preconditioning matrices led to much faster convergence rates. Furthermore, if 
the requirement of monotonic convergence of the iterates to the bound is dispensed with, there 
exists several algorithms with even faster convergence rates. The work also presents a way of 
approximating the inverse of singular FIMs. 

In this paper, we show that the algorithms proposed in prior work are special instances of a more 
general framework related to solving a \textit{quadratic matrix program} \cite{Beck07QMP}, a 
generalization of the well-known \textit{quadratic program}, a convex optimization problem 
\cite{Boyd04Opt}. The reformulation provides a framework to develop methods for fast computation of the
CRB, and explore various computational trade-offs. Consequently, the vast literature in convex 
optimization can be exploited. Our formulation enables us to extend to the cases when the parameters 
are constrained \cite{Gorman90CRB,Stoica98CRB} and when the Fisher information matrix is
singular, with ease. The work done here may be of independent interest to other areas when a
similar motivation is required. We then apply these methods on an application related to the design
of a specific data streaming algorithm for measuring flows through a router. By doing so, we are
able to compare the performance of several constrained optimization methods.

We denote all vectors and matrices with lower case and upper case bold 
letters respectively. Random variables are italicized upper case letters. Sets are denoted with 
upper case calligraphic font. We work entirely in the real space $\Real$. $\Spd$ and $\Spsd$ denote
the set of real-valued, symmetric positive definite and positive semidefinite matrices of size $n$.
The matrix $\diag(\bx)$ is a diagonal matrix with elements of $\bx$ on its diagonals. 
$\tr(\bA)$ and $\rank(\bA)$ denote the trace and rank of a matrix $\bA$ respectively. The 
eigenvalues of $\bA$ are denoted by $\lambda_1(\bA) \ge \lambda_2(\bA) \ge \ldots \ge 
\lambda_n(\bA)$, ordered from maximum to minimum. Vector $\bEu{i}$ denotes the $i$-th canonical 
Euclidean basis in $\Real^n$. $\|\bx\|_2$ and $\|\bX\|_F$ denotes the Euclidean and Frobenius norm 
of vector $\bx$ and matrix $\bX$ respectively. Other notation will be defined when needed.

\section{Preliminaries}
\label{sec:prelim}

\subsection{Fisher information}

Let the real, non-random parameter vector be denoted by $\vth = \lbrack \theta_1, \theta_2,
\ldots, \theta_n \rbrack^\T$. The parameter $\vth \in \bTheta$, where $\bTheta \subseteq \Real^n$
is an open set. Let $\{P_{\vth}\}_{\vth \in \bTheta}$ be a family of probability measures for a 
certain random variable $\iY$ taking values in set $\mathcal{Y}$. Assume that $P_{\vth}$ is 
absolutely continuous with respect to a dominating measure $\mu$ for each $\vth \in 
\bTheta$. Thus, for each $\vth$ there exists a density function $f(\by;\vth) = d P_{\vth}/d\mu$
for $\iY$. We define the expectation $\E_{\vth}\lbrack \iY \rbrack = \int \by\,dP_{\vth}$ whenever
$\int |\by|\,dP_{\vth}$ is finite.

We assume that the family of densities $\{f_{\iY}(\by;\vth)\}_{\vth \in \bTheta}$ is regular, 
i.e.~satisfying the following three conditions:
(1) $f_{\iY}(\by;\vth)$ is continuous on $\bTheta$ for $\mu$-almost all $\by$,
(2) the log-likelihood $\log f_{\iY}(\by;\vth)$ is mean-square differentiable in $\vth$, and 
(3) $\nabla_{\vth} \log f_{\iY}(\by;\vth)$ is mean-square continuous in $\vth$.
These conditions ensure the existence of the FIM
\be
\bJ := \E_{\vth}\lbrack \nabla_{\vth} \log f_{\iY}(\by;\vth)\rbrack \lbrack \nabla^\T_{\vth} \log 
f_{\iY}(\by;\vth)\rbrack,
\label{eq:fisher_information}
\ee
which is an $n \times n$ positive semidefinite matrix and is finite.
With the assumption of the existence, continuity in $\vth$ and absolute integrability in $\iY$ of 
the mixed partial differential operators $(\partial^2/\partial \theta_i \partial \theta_j) 
f_{\iY}(\by;\vth)$, $i,j = 1,2,\ldots,n$, the FIM becomes equivalent to the Hessian of
the mean of the curvature of $\log f_{\iY}(\by;\vth)$, $\bJ = -\E_{\vth} \nabla^2_{\vth} \log
f_{\iY}(\by;\vth)$.

\subsection{Cram\'er-Rao Bound}

The importance of the Fisher information is its relation to the Cram\'er-Rao bound (CRB). The CRB is
a lower bound on the covariance matrix of any unbiased estimator of the parameter $\vth$. For any 
unbiased estimator $\veth(\by)$ on observations $\by$, the relation is given by
\be
\E\lbrack (\veth(\by) -\vth)(\veth(\by)-\vth)^\T \rbrack \ge \bJ^{-1}.
\label{eq:crb}
\ee
In principle, it is possible to compute submatrices of the CRB by partitioning the Fisher
information matrix into blocks, and then apply the matrix inversion lemma \cite{Harville97MtxStat}. 
As reported in \cite{Hero94CRBComp}, methods such as sequential partitioning \cite{Horn85Matrix}, 
Cholesky and Gaussian elimination require $O(n^3)$ flops. These methods have a high number of
flops even if we are concerned with a small submatrix, for e.g.~the covariance of just $m \ll n$
parameters, motivating our work.

\section{Formulation and Algorithms}
\label{sec:algo}

As a start, we assume a nonsingular Fisher information matrix $\bJ$. We now consider the 
optimization problem
\begin{eqnarray}
\underset{\bx \in \Real^n}{\text{min}} & \frac{1}{2}\bx^\T \bJ \bx - \bb^\T\bx ,
\label{eq:qp}
\end{eqnarray}
an example of a quadratic program \cite{Boyd04Opt}. Let $F(\bx) := \frac{1}{2}\bx^\T \bJ \bx -
\bb^\T\bx$. In this case, the optimization problem is strictly convex and possesses a unique
optimal. The unique optimal solution to this problem is $\bx^\star = \bJ^{-1} \bb$. The
generalization of the above is the \textit{quadratic matrix program},
\begin{eqnarray}
\underset{\bX \in \Real^{m\times n}}{\text{min}} & \frac{1}{2} \tr(\bX^\T \bJ \bX) - 
\tr(\bB^\T\bX) .
\label{eq:qmp}
\end{eqnarray}
Any feasible solution to \eqref{eq:qmp} is a valid lower bound on the covariance of 
the unbiased estimate of $\vth$, and the tightest lower bound (the CRB) is the global minimum to 
\eqref{eq:qmp} \cite{Stoica01SingularFIM}. Matrix $\bB$ focuses the computation on a submatrix of 
the CRB. The special case $\bB = \id{n}$ is equivalent to performing the full inverse of $\bJ$, 
while setting $\bB = \bEu{k}$ enables computation of the CRB of just a single $\theta_k$, 
$k \in \{1,2,\cdots,n\}$. The optimal solution to this problem is $\bX^\star = \bJ^{-1} \bB$ (see 
Appendix). Based on this, if an algorithm searches for the minimum of the optimization problems 
\eqref{eq:qp} or \eqref{eq:qmp}, it effectively computes $\bb^\T \bJ^{-1} \bb$ and $\bB^\T 
\bJ^{-1} \bB$ respectively, essentially computing the CRB.

\subsection{Majorization-Minimization (MM) methods}

The optimization problems above can be solved via the majorization-minimization (MM) method, which
generalizes the Expectation-Maximization (EM) method (see \cite{Lange04MM} for 
a tutorial). Thus, the algorithm in \cite{Hero94CRBComp} is a special instance of MM. We show that 
the recursive bounds of \cite{Hero94CRBComp} are just the consequence of a special choice. 

We first consider the vector case. Define
\be
G(\bx;\bxi{k}) := \frac{1}{2}\bx^\T \bJ \bx - \bb^\T\bx + Q(\bx;\bxi{k}).
\label{eq:majorizer}
\ee
The function $Q(\bx;\bxi{k})$ must be chosen so that $G(\bx;\bxi{k})$ \textit{majorizes} $F(\bx)$. 
There are two properties to satisfy:
(1) $G(\bx;\bxi{k}) \ge F(\bx)$ for all $\bx$, and
(2) $G(\bxi{k};\bxi{k}) = F(\bxi{k})$.
These requirements ensure that $G(\bx;\bxi{k})$ lies above the surface of $F(\bx)$ and is tangent at 
the point $\bx = \bxi{k}$ \cite{Lange04MM}. The function $G(\bx;\bxi{k})$ is referred to as a 
\textit{surrogate function}. By these properties, MM-based algorithms converge to the CRB 
\textit{monotonically}.

For example, suppose we have a matrix $\bP \in \Spd$ and $\bP \ge \bJ$ in the positive 
semidefinite sense, then
\be
Q(\bx;\bxi{k}) := \frac{1}{2}(\bx - \bxi{k})^\T (\bP - \bJ) (\bx - \bxi{k})
\label{eq:q_function}
\ee
is a popular choice. Minimizing \eqref{eq:majorizer} with the choice \eqref{eq:q_function} w.r.t.~$\bx$ results in a 
closed--form solution 
\begin{align}
\label{eq:jacobi_iteration}
\bxi{k+1} &= (\id{n} - \bP^{-1}\bJ) \bxi{k} + \bP^{-1} \bb
= \bxi{k} + \bP^{-1} (\bb - \bJ \bxi{k}).
\end{align}
The above is simply a \textit{Jacobi iteration}, with a preconditioner $\bP$ 
\cite{Watkins02MtxComp}. Typically, $\bP$ is chosen to be diagonal or near diagonal, as this
facillitates simple computation of $\bP^{-1}$. Setting $\bP$ to the Fisher information matrix of the complete 
data space would yield the algorithm in \cite{Hero94CRBComp}. For the matrix case, 
$G(\bX;\bXi{k}) := \frac{1}{2}\tr\Big(\bX^\T \bJ \bX - \bb^\T\bX\Big) + Q(\bX;\bXi{k})$,
we have the choice $Q(\bX;\bXi{k}) := \frac{1}{2}\tr\Big((\bX - \bXi{k})^\T (\bP - \bJ) (\bX - 
\bXi{k})\Big)$, to obtain a Jacobi iteration.

The convergence rate for this particular choice is governed by the spectral radius $\rho(\id{n} 
- \bP^{-1}\bJ)$, which is the maximum magnitude eigenvalue of the matrix. Exponential convergence 
to the CRB is achieved by ensuring that $\rho(\id{n} - \bP^{-1}\bJ) < 1$ is as small as
possible. It is for this reason $\rho(\id{n} - \bP^{-1}\bJ)$ is also referred to as the 
\textit{root convergence factor} \cite{Watkins02MtxComp}, which measures the asymptotic convergence rate. 

The power of MM lies in the great freedom of choice when designing $Q(\bX;\bXi{k})$. For fast convergence, 
one needs to choose a $Q(\bX;\bXi{k})$ that well-approximates the quadratic objective around $\bXi{k}$. 
Second, $Q(\bX;\bXi{k})$ is chosen in a way that it does not depend on quantities we desire, such 
as $\bJ^{-1}$ or is computationally expensive, for instance, a dense $\bP$. These trade-offs 
make the algorithm design more of an art than science.

\subsection{Gradient Descent (GD) methods}

Gradient descent methods rely on minimizing the function along particular search directions. At 
each iteration, two crucial elements are required: the search direction $\bdi{k}$, and the size of
the step $\iomega{k}$. Algorithm \ref{alg:grad_descent} presents a generic outline of gradient 
descent methods.

\begin{algorithm}[h]
\caption{Generic implementation of gradient descent}
\begin{algorithmic}[1]
\REQUIRE $\epsilon$, error threshold 
\STATE $\bxi{0} \leftarrow \bx_{\text{init}}$

\WHILE{$\|F(\bxi{k}) - F(\bxi{k-1})\|_{2} \ge \epsilon$}
\STATE $\iomega{k+1} \leftarrow \arg \min_{\omega} F(\bxi{k}+\omega \bdi{k})$ \COMMENT{Exact line search}
\STATE $\bxi{k+1} \leftarrow \bxi{k} + \iomega{k+1} \bdi{k}$
\STATE $\bdi{k+1} \leftarrow \Delta(\bxi{k+1})$
\ENDWHILE
\end{algorithmic}
\label{alg:grad_descent}
\end{algorithm}

Gradient methods depend on the evaluation of a function $\Delta(\bx)$ which determines the search
direction. For example, in classical gradient descent, this is simply gradient of $F(\bx)$ at each
iterate, $\Delta(\bx) = -\nabla_{\bx} F(\bx) = \bb-\bJ\bx$. 

Exact line searches, however, can be computationally expensive. With a fixed choice of $\omega$ 
such that $\omega < 2/\lambda_n(\bJ)$, the algorithm uses an \textit{inexact} line search, equivalent to the 
\textit{Richardson iteration} \cite{Watkins02MtxComp}. The Gauss-Seidel (GS) method performs
$\Delta(\bxi{k}) = -\bEu{(k \mod n) + 1}$, $i=1,2,\ldots,n$ \cite{Watkins02MtxComp}. Conjugate and 
preconditioned conjugate gradient algorithms \cite{Watkins02MtxComp} also belong to this class,
where the search directions are constructed by performing a Gram-Schmidt procedure. Hence, some of the recursive 
algorithms presented in \cite{Hero96CRBRecursive} are all instances of gradient descent methods. Unlike the algorithms 
presented in the previous section, these algorithms generally have \textit{non-monotonic} convergence to the CRB.
We particularly advocate preconditioned conjugate gradient algorithms for their fast convergence, 
requiring only simple line searches, and shown to have excellent performance in Section \ref{sec:application}.

The Newton-Raphson descent method is unusable here, since it requires the inverse of the Hessian 
of the objective function, which is $\bJ^{-1}$, of which we are avoiding its direct computation.
For this reason, it is much better to use methods with simple line searches with low memory 
requirements.

The gradient search method can be applied to quadratic matrix programs. Some adaptation is needed,
however, such as reformulating the problem to a suitable vectorized form (see \cite{Beck07QMP}). 

\subsection{Extension to Singular Fisher Information}

Suppose now $\bJ$ is singular. Such matrices arise in some areas, such as blind channel 
estimation \cite{BlindEst98} and positron emission tomography \cite{Hero96BVtradeoff}. The properties 
of singular $\bJ$ were explored in \cite{LiuBrown93Singular,Stoica01SingularFIM}. 

The approach taken in \cite{Hero96CRBRecursive} was to add a perturbation to $\bJ$ in order to make
it nonsingular, and then compute its inverse via recursive algorithms described above. This approach
only yields an approximation to the CRB, with increased computational complexity. Instead, we take a 
completely different, more efficient, route.

Assuming $\bb \in \range(\bJ)$, consider the optimization problem for the vector case,
\begin{eqnarray}
\underset{\bx \in \Real^n}{\text{min}} & \frac{1}{2}\|\bb - \bJ\bx \|_2.
\label{eq:norm_minimization}
\end{eqnarray}
The optimization problem is convex and the solution is simply the minimum norm solution $\bx^\star 
= \bJ^{+} \bb$, where $\bJ^{+}$ denotes the Moore-Penrose pseudoinverse \cite{Harville97MtxStat},
which is unique. Thus, we can solve for the CRB without having to resort to the approach
taken by \cite{Hero96CRBRecursive}.

The generalization of \eqref{eq:norm_minimization} is
\begin{eqnarray}
\underset{\bx \in \Real^n}{\text{min}} & \frac{1}{2}\|\bB - \bJ\bX \|_F,
\label{eq:norm_minimization_mtx}
\end{eqnarray}
assuming the column space of $\bB$ is in $\range(\bJ)$. Then, the minimum norm solution is 
$\bX^\star = \bJ^{+} \bB$. The optimization problems here can be solved via the MM or GD methods.
As an example, using the MM method in the vector case, and choosing \eqref{eq:q_function} results 
in  
%\be
$\bxi{k+1} = \bxi{k} + \bP^{-1} \bJ(\bb - \bJ \bxi{k})$.
%\label{eq:landweber}
%\ee
A variation where $\bP = \nu \id{n}$ with $\nu \ge \lambda_1(\bJ^2)$ is the well-known 
\textit{Landweber iteration} \cite{Strand74Landweber}. The technique also applies to problem
\eqref{eq:norm_minimization_mtx}.

\subsection{Extension to Constrained Fisher Information}

Certain constraints provide additional information, resulting in the reduction of estimator 
variance. A direct way of deriving the constrained Fisher information is to recompute the Fisher 
information with a new vector parameter $\boldsymbol{\gamma}$ such that constraints are
incorporated in $\boldsymbol{\gamma}$. Unfortunately, it generally requires a nontrivial
alteration of the p.d.f.'s dependence on $\boldsymbol{\gamma}$ instead. Often, the approach is
analytically intractable or numerically complex. The papers
\cite{Gorman90CRB,Stoica98CRB} were motivated by this problem and provide analytic
formulae to compute the constrained Fisher information matrix without reparameterization.

In the following, it is enough to assume the unconstrained Fisher information matrix $\bJ \in
\Spsd$. It has been shown that inequality constraints do not affect the CRB \cite{Gorman90CRB},
thus, we focus entirely on equality constraints. Assume there are $p$ consistent and nonredundant
equality constraints on $\vth$, i.e.~$h(\vth) = \mathbf{0}$. Let $\bH \in \Real^{n \times p}$
denote the gradient of constraints $h(\vth)$. By the nonredundancy of the constraints,
$\rank(\bH) = p$. Furthermore, let $\bU \in \Real^{n \times (n-p)}$ be a matrix whose column space
is the orthogonal basis of the cokernel of $\bH$, i.e.~$\bH^\T\bU = \mathbf{0}$, and $\bU^\T
\bU = \id{n-p}$. 

Assuming that $\bU^\T \bJ \bU$ is nonsingular, it has been shown that the CRB is simply 
\cite{Stoica98CRB}
\be
\I^+ = \bU(\bU^\T \bJ \bU)^{-1}\bU^\T.
\label{eq:constrained_fisher}
\ee
If $\bJ$ is nonsingular, the above can be rewritten as $\I^{+} = \bJ^{-1} - 
\bJ^{-1}\bH(\bH^\T\bJ^{-1}\bH)^{+}\bH^\T \bJ^{-1}$, which is equivalent to the bound derived in
\cite{Gorman90CRB}, by choosing $\bU = \id{n} - \bJ^{-1}\bH(\bH^\T\bJ^{-1}\bH)^{+}\bH^\T$.

The algorithms proposed in \cite{Hero94CRBComp,Hero96CRBRecursive} no longer apply here. It is 
also hard to see how the recursive algorithms can be extended to account for parameter
constraints. It turns out constraints can be incorporated into our framework naturally.
 
The solution to the optimization problem (proof in Appendix \ref{app:constrained})
\begin{eqnarray}
\label{eq:constrained_qp}
\underset{\bX \in \Real^{m\times n}}{\text{min}} & \frac{1}{2} \tr(\bX^\T \bJ \bX) - 
\tr(\bB^\T\bX) \\
\nonumber
\text{subject to} & \bH^\T \bX = \mathbf{0}.
\end{eqnarray}
is simply
\be
\bX^\star = \bU(\bU^\T \bJ \bU)^{-1}\bU^\T\bB = \I^{+} \bB.
\label{eq:opt_constrained}
\ee
From this, we have computed submatrices of the constrained CRB, extending the work in
\cite{Hero94CRBComp}. The general shift to a quadratic matrix program instead enables us to consider
constraints naturally. If $\bJ$ is nonsingular, then equation \eqref{eq:opt_constrained} is equivalent to 
\be
\bX^\star = \bJ^{-1}\bB - \bJ^{-1}\bH(\bH^\T\bJ^{-1}\bH)^{+}\bH^\T \bJ^{-1}\bB,
\label{eq:opt_constrained_ns}
\ee
in agreement with the above.

The algorithms discussed previously require some modifications to account for constraints. MM 
methods are still applicable by ensuring constraints are built into the recursion. GD methods such
as the preconditioned conjugate gradients algorithm can be adapted with constraints (for
e.g.~\cite{Coleman94PCG}). We test some of these methods below.

\def\Opt				{\centering\includegraphics[width=\twoup]{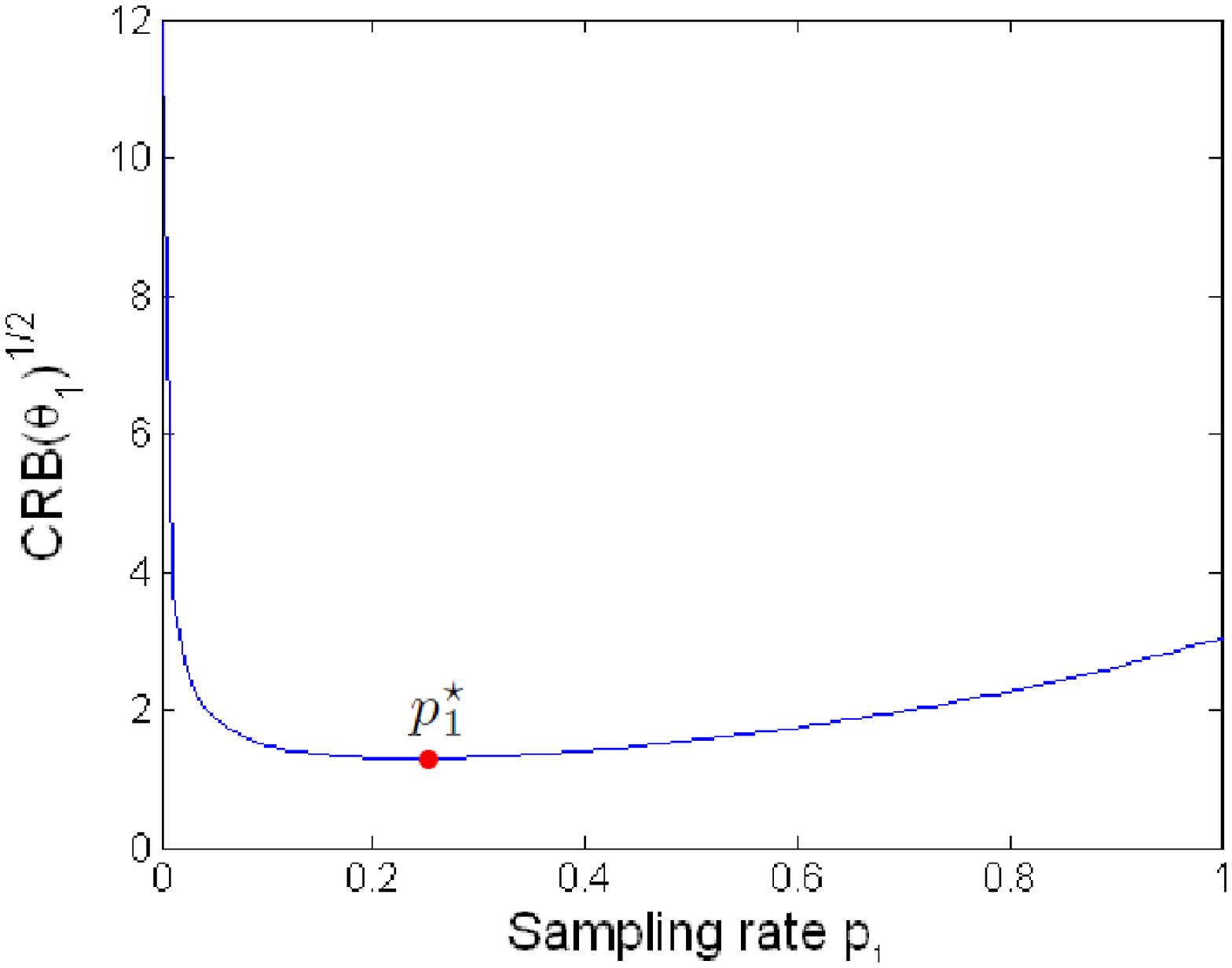}}
\def\Monotonic	{\centering\includegraphics[width=\twoup]{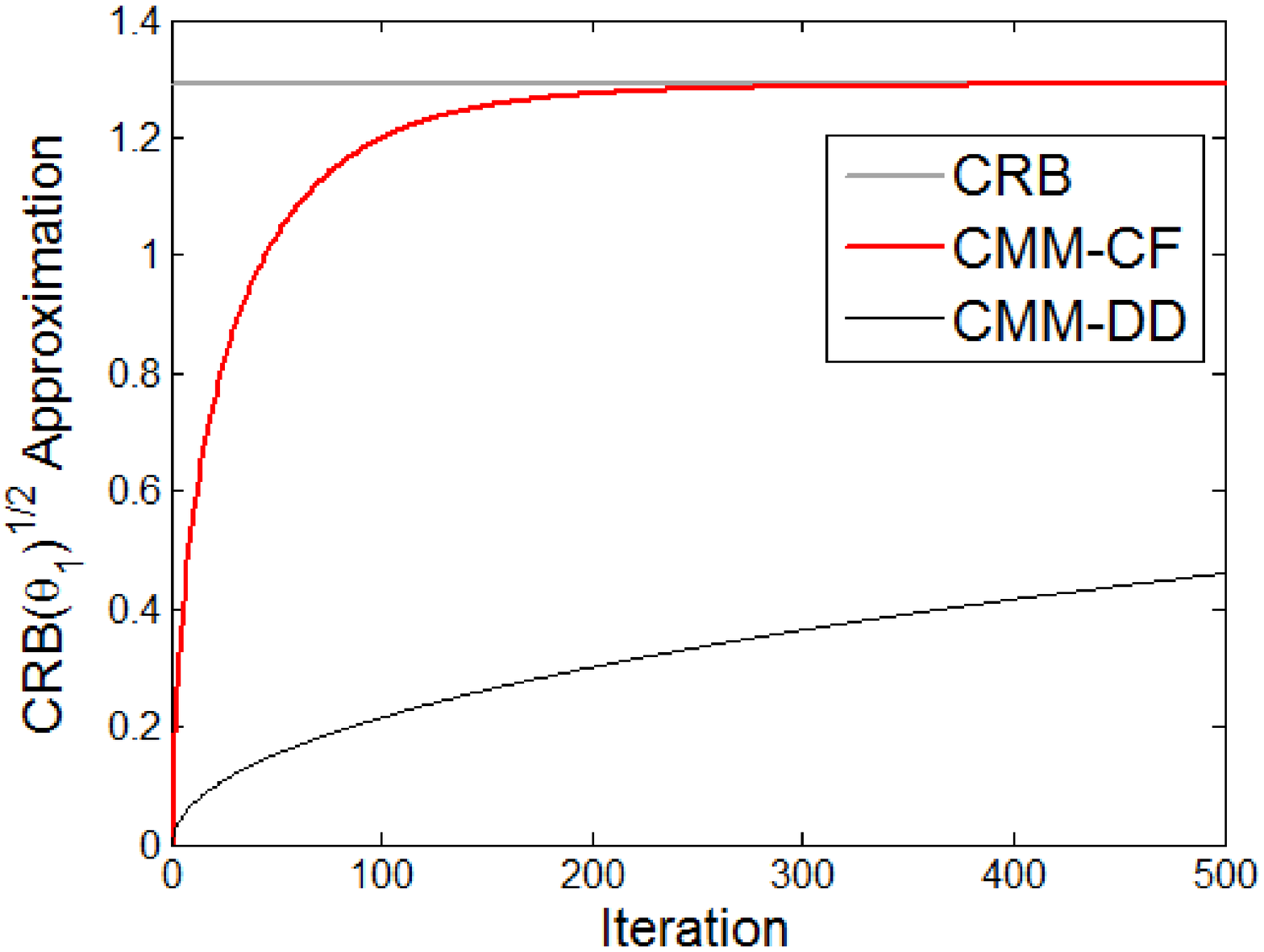}}
\def\Trajectory	{\centering\includegraphics[width=\twoup]{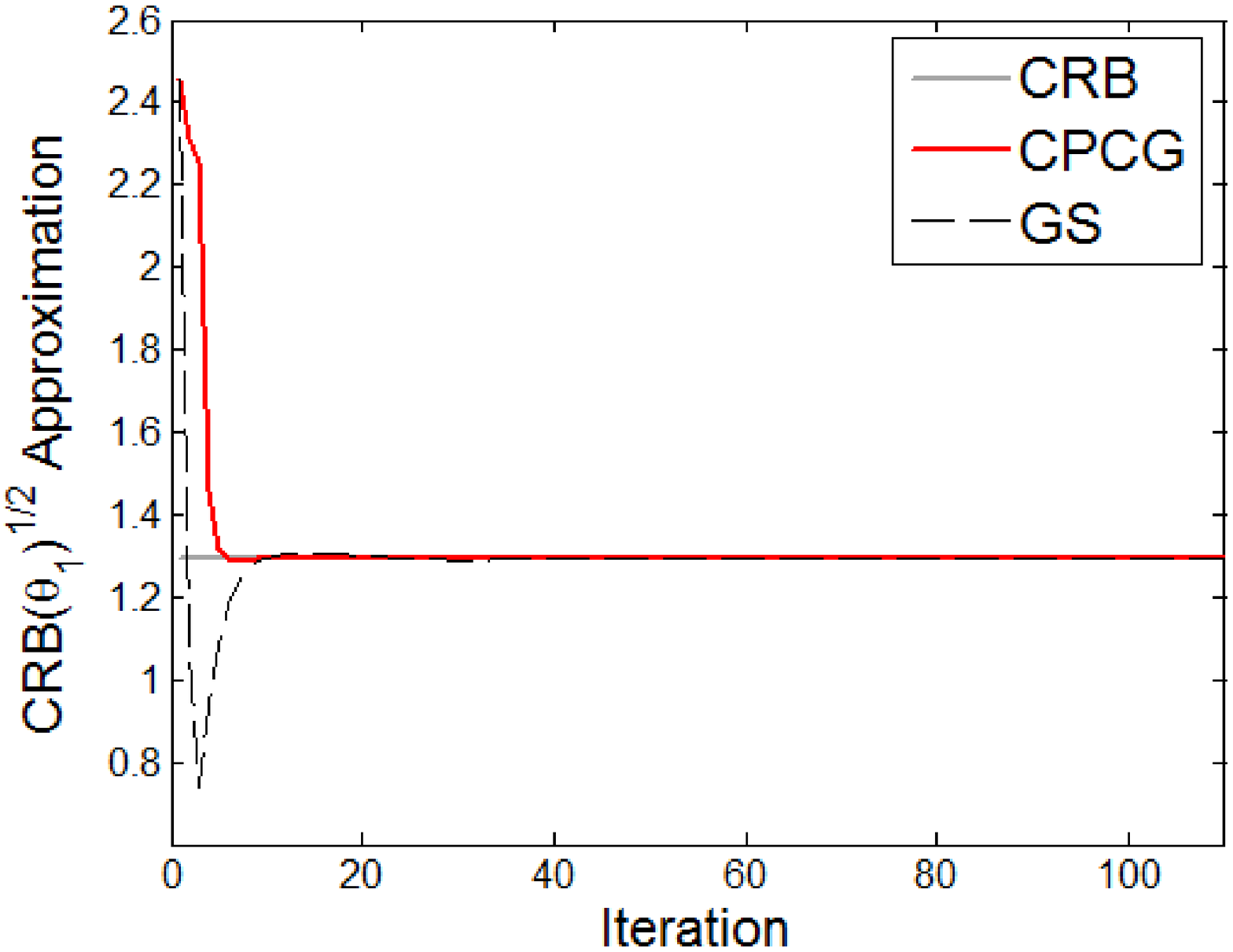}}

\section{Application}
\label{sec:application}

In this section, due to space limitations, we perform numerical experiments to test the efficiency 
of the algorithms on only one example. The application involves the optimization of a data 
streaming algorithm for the measurement of flows on networks, where the parameters $\vth$ are 
subject to constraints.

\subsection{Data Streaming Algorithm Optimization}

A \textit{flow} is a series of packets with a common key, such as the source and destination
Internet Protocol address. The \textit{flow size} is defined as the number of packets it contains.
We are interested in the flow size distribution $\vth = \lbrack \theta_1,\theta_2, 
\ldots,\theta_n\rbrack^\T$ in a measurement interval of $T$ seconds. Each $\theta_k$ denotes the
proportion of flows of size $k$, with $n$ being the largest flow size. By definition, $\sum_{k=1}^n
\theta_k = 1, \theta_k > 0, \forall k$ (the strict inequality of the latter constraint to ensure no
bias issues arise, see \cite{Tune11DSjournal}). The gradient of the equality constraint is $\onen$.

Data streaming algorithms are used for measuring flows on core networks, since the huge volume and
speed of flows imposes strict memory and processing requirements. The advantage of these algorithms
is the small amount of memory required, at the expense of introducing some error when recovering flow
traffic statistics. 

The data streaming algorithm we consider is the flow sampling--sketcher (FSS). FSS has an array of
$A$ counters. Every incoming flow is selected with i.i.d.~probability $p$ and dropped with
probability $q$. Sampling is performed via the use of a sampling hash function
$h_s(x)$ with full range $R$, which acts on a flow key $x$, configured such that a packet is
accepted if $h_s(x)\le p R$. The deterministic nature of the hash function ensures that packets 
belonging to a sampled flow will be always sampled and vice versa. For packets of sampled 
flows, another hash function $h_c(x)$ with range $A$ generates an index, ensuring that the same
counter is incremented by packets from the same flow. The counter with the corresponding index is
incremented once per packet. Note that several flows can be mapped to the same counter, 
resulting in \textit{collisions}. Once the measurement interval is over, the flow size 
distribution is recovered by employing an EM algorithm. FSS is practically implementable in routers.
The schemes in \cite{Kumar04Counter,Ribeiro08Minimalist} are closest in spirit to FSS.

Let $\nf$ be the total number of flows in the measurement interval and $\alpha' = p\nf/A$ denote 
the average number of flows in a counter. Assuming fixed $A$ (i.e.~fixed memory allocation),
the latter has a direct impact on estimation quality, as $\alpha'$ controls the flow collisions in
the counters. Sampling with low $p$ would increase variance due to missing flows, 
while high $p$ results in many flows mapping to the same counter, increasing ambiguity due to 
collisions. Due to the dependence between a flow size $k$ on flows smaller than it in each 
counter, different optimal sampling rates $p^{\star}_{k}$ minimize the
estimator variance for each $\theta_k$. The objective is to find $p^{\star}_{k}$ for a particular
target flow size $k$, for e.g.~$k=1$ which is especially important for detecting network attacks.

We use the Poisson approximation to compute the counter array load distribution, $\bc = 
\lbrack c_0(\vth),c_1(\vth),\cdots \rbrack^\T$. The generating function of the load distribution is ($|s| < 1$ for convergence)
\be
C^*(s;\vth) = \prod_{k=1}^n e^{\alpha' \theta_k (s^k - 1)} = e^{-\alpha'}\cdot  e^{\sum_{k=1}^n \alpha' \theta_k s^k},
\label{eq:load_gf}
\ee
essentially a convolution of $n$ weighted Poisson mass functions (see \cite{Tune11Skampling}). The distribution $\bc$
is obtained from the coefficients of the polynomial expansion of $C^*(s;\vth)$, easily computed via the Fast Fourier Transform (FFT).
Examples include $c_0(\vth) = e^{-\alpha'},\ c_1(\vth) = \alpha'\theta_1e^{-\alpha'},\ c_2(\vth) = (\alpha'\theta_2 + \tfrac{\alpha'^2 \theta^2_1}{2!})
e^{-\alpha'}$. Let $\bW = \diag(c^{-1}_1(\vth),c^{-1}_2(\vth),
\cdots)$. The unconstrained Fisher information is
\be
\bJ(p) = q \onen \onen^\T + p \alpha' \bG^\T \bW \bG, 
\label{eq:skampler_uncon_alt}
\ee
where the matrix $\bG$ is a quasi-Toeplitz matrix with generating sequence $\mathbf{c}_{\vth}$, and
$\bJ(p)$ is positive definite $\forall p$  \cite{Tune11Skampling}. The matrix is dense and is challenging to compute.
Computing the constrained Fisher information $\I^{+}(p)$ is even more difficult
(c.f.~\eqref{eq:constrained_fisher}). Since we only need to compute individual diagonal entries of
$\I^{+}(p)$, as each $k$-th diagonal is the CRB of the estimator variance of $\theta_k$, and
defining 
\begin{eqnarray}
\label{eq:opt_variance}
g(\bx^\star,p) &= \min_{\bx \in \Real^n} &  \frac{1}{2}\bx^\T \bJ(p) \bx - \bEu{k}^\T\bx\\
\nonumber
& \hspace{5mm} \text{subject to} & \onen^\T \bx = \mathbf{0},
\end{eqnarray}
we can use, assuming unimodality of $g(\bx^\star,p)$ w.r.t.~$p$, (see Appendix for justification)
\begin{eqnarray}
\label{eq:opt_sketch}
p^{\star}_{k} &= \arg \max_{p \in (0,1\rbrack}\hspace{5mm} g(\bx^\star,p).
\end{eqnarray}
For fixed $p$, since problem \eqref{eq:opt_variance} is equivalent to \eqref{eq:constrained_qp} , we can use 
efficient optimization algorithms to solve it, avoiding full inversion. Problem 
\eqref{eq:opt_sketch} can be solved by a golden section search \cite{ArtNumerical07}. The generality of our approach allows 
us to, for e.g.~compute $p^\star_{\lbrack k, \ell\rbrack}$, the optimal sampling rate that minimizes
the joint variance of $\theta_k$ to $\theta_{\ell}$, achieved by replacing $\bEu{k}$ with a matrix
$\bB$ which is zero everywhere except for the matrix $\id{\ell-k+1}$ at the $\ell$-th position, 
and use \eqref{eq:constrained_qp} in place of problem \eqref{eq:opt_variance}. 

\begin{figure}[t]
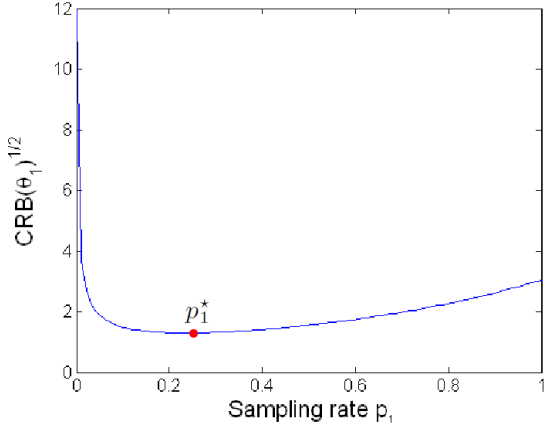

\Opt 
\caption{Sampling rate $p$ against the square root of CRB of $\theta_1$ for the distribution
Abilene-III, truncated at $n = 2,000$, and $\alpha = 4$. The optimal sampling rate $p^{\star}_1 =
0.2342$, denoted by the dot on the curve.}
\label{fig:optimal_abi3}
\end{figure}

\subsection{Numerical Results}

Our focus is the computation of \eqref{eq:opt_variance} using the various methods discussed
earlier. We tested the algorithms on the important case of $k = 1$. The distribution used is a
truncated version of the Abilene-III \cite{AbileneIII}, truncated to $n = 2,000$ packets to
the satisfy the parameter constraints. Here, $\alpha = 4$, $p^{\star}_{1} = 0.2342$ and 
tolerance for all iterative algorithms is within $\epsilon = 10^{-6}$ of the true CRB. For reference, 
$\lbrack \I^{+}(p^{\star}_{1})\rbrack_{11} = 1.67005$. While it is unknown if the sampling rate-CRB 
curve is strictly convex for all $\theta_k$, in this case, it is (see Figure \ref{fig:optimal_abi3}(a)). 
We omit dependence on $p$ in the following since $p = p^{\star}_{1}$.

In practice, $\bc$ is truncated up to a sufficiently large number of terms $K$ and computed using Fast
Fourier transforms. In what follows, we assume $\bc$ has been computed. Define $\mathbf{J}_{\vth,M}$ 
to be the Fisher information computed with $\bc$ up to $M$ terms. Then, $K$ is chosen as the value 
when $\|\mathbf{J}_{\vth,K} - \mathbf{J}_{\vth,K-1}\|_F < \delta$, i.e.~a preset tolerance 
$\delta > 0$. With the value of $K$ terms, it takes $n^2(K+1)$ flops to construct $\bJ$. In our 
case, $K = 10,000$. We assume that $\bJ$  is computed and stored upfront for all methods. In our case, this is cheaper than 
recomputing vector-matrix products $\bJ \bx$, due to the complexity of $\bG$, at the expense of 
higher memory storage.

If we perform full inversion, it takes $n^3/3$ flops via Cholesky methods, followed by $3n^2$ flops 
to construct the term $\bJ^{-1}\onen(\onen^\T\bJ^{-1}\onen)^{-1}\onen^\T \bJ^{-1}$. Thus, it takes a 
total of $n^3/3 + 3n^2$ flops, and with $n = 2,000$, it takes 2.68 Gflops. In contrast, for recursive 
methods, each iteration requires $(n+1)^2$ flops, with the additional requirement to account for 
constraints. Depending on the method, additional operations might be required such as computing the 
diagonal preconditioner, which would require about 9$n$ flops (see \cite{Hero96CRBRecursive}). 
Generally, $O(n^2)$ flops per iteration are required for the following methods.

We compare two classes: Constrained Majorization-Minimization (CMM) and Constrained Preconditioned
Conjugate Gradient (CPCG). For CMM, we have CMM-CF where the Fisher information of the complete data 
space, $\mathbf{\bar J}_{\vth} = \alpha\,\diag(\theta^{-1}_1, \theta^{-1}_2,\cdots, \theta_n^{-1})$ 
was used as the preconditioning matrix. CMM-DD instead uses the first order diagonally dominant 
matrix of $\bJ$ (see \cite{Hero96CRBRecursive} for more details). The recursion step for CMM was
derived using Lagrangian multipliers to account for constraints when minimizing \eqref{eq:majorizer}. 
For CPCG, the preconditioner matrix used is $\mathbf{\bar J}_{\vth}$. 
 
%We also have CMM-R, which is equivalent to a Richardson
%iteration, using a preconditioning matrix $\nu \id{n}$, where $\nu =
%\tfrac{\lambda_{1}(\bJ)+\lambda_{n}(\bJ)}{2}$ is the optimal value \cite[p.~552]{Watkins02MtxComp}.

We also tested Gradient Projection (GP), which is the standard GD algorithm using $\mathbf{\bar 
J}_{\vth}$ as preconditioner, but accounts for constraints and uses exact line searches. GP, however, 
diverged for all iterations. Even without a preconditioner, the results remain the same. We omit its 
results and explain its poor performance later on.
 
All algorithms were initialized with the same initial point. The \textit{breakeven threshold}, i.e. 
the number of iterations before all methods would lose computational advantage to a direct evaluation
of the CRB is 667 iterations. Table \ref{table:alg_performance} presents the comparison between
different algorithms. The second column lists the root convergence factor of each algorithm. The root
convergence factor, $\rho$ is defined differently for each method. For CMM, refer to Section
\ref{sec:algo}.For CPCG, it is $\rho = \tfrac{\sqrt{\kappa}-1}{\sqrt{\kappa}+1}$, where $\kappa =
\tfrac{\lambda_1(\mathbf{\bar J}^{-1}_{\vth}\bJ)}{\lambda_n(\mathbf{\bar J}^{-1}_{\vth}\bJ)}$
\cite{Watkins02MtxComp}. The third and fourth columns lists the number of iterations required for the
result from the algorithms to be $5\%$ and $0.5\%$ respectively, tolerance of the CRB. The fifth
column denotes the number of iterations before the algorithm reaches within tolerance $\epsilon$ of
the CRB.  

\begin{figure}[ht]
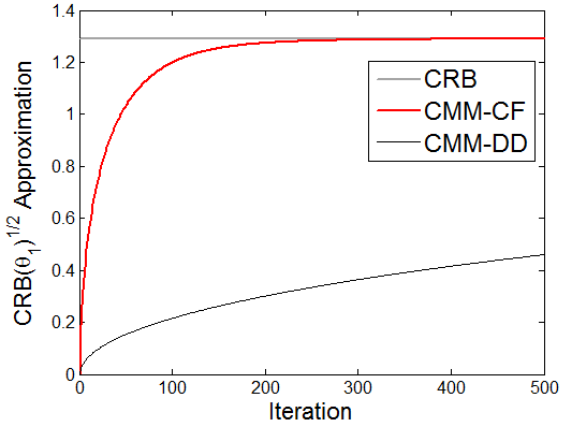

\Monotonic
\caption{Trajectory of CMM-CF and CMM-DD when computing the square root of CRB of $\theta_1$ for the 
distribution Abilene-III, truncated at $n = 2,000$, and $\alpha = 4$, shown here up to 500 
iterations. Tolerance $\epsilon = 10^{-6}$. CMM-CF converged in 8,487 iterations, while CMM-DD 
converged in 69,026 iterations. Note the monotonic convergence of both algorithms to the true CRB.}
\label{fig:monotonic}
\end{figure}

While all CMM algorithms converge monotonically to the bound, they are extremely slow. The monotonic
convergence of both algorithms can be seen in Figure \ref{fig:monotonic}. Clearly, CMM-CF has a faster
convergence rate compared to CMM-DD. CPCG performs the best; however, its iterates have non-monotonic 
convergence. CMM and GP perform badly due to the small condition number of $\bJ$, which is
$2.73\times 10^6$. In particular, for GP, the projected descent steps move in a circular
trajectory. Note that for all methods, the root convergence factor is a good predictor of the total
number of iterations needed for convergence, but is not predictive of the number of iterations needed
to be within $5\%$ and $0.5\%$ of the bound. Furthermore, only CPCG possesses some robustness with
respect to the selection of the initial point. The other algorithms have a strong dependence on the
initial point, and may have poor performance with a bad initial point choice. Finally, CPCG is the 
only algorithm that converges within the breakeven threshold.

\begin{figure}[t]
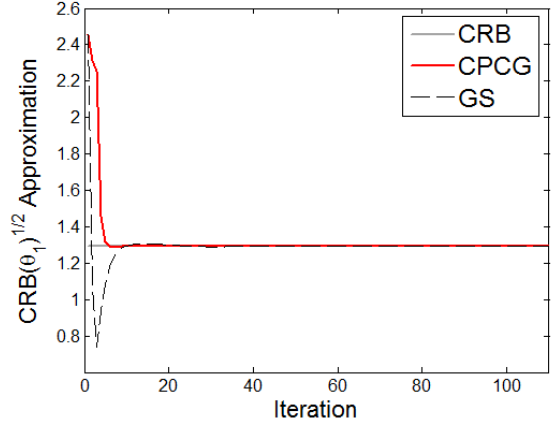

\Trajectory
\caption{Trajectory of the GS and CPCG when computing the square root of CRB of $\theta_1$ for the 
distribution Abilene-III, truncated at $n = 2,000$, and $\alpha = 4$. Tolerance $\epsilon = 10^{-6}$. 
GS converged in 110 iterations, while CPCG converged in 64 iterations.}
\label{fig:trajectory}
\end{figure}

We also compare CPCG against a straightforward way of evaluation using the GS method. 
We use GS to evaluate two quantities separately: $\bJ^{-1}\bEu{1}$ and $\bJ^{-1}\onen$, and then use 
\eqref{eq:opt_constrained_ns} to evaluate $\bx^\star$. The trajectory of this method was compared 
with the trajectory of CPCG in Figure \ref{fig:trajectory}. Initialization for GS requires two 
initial points for evaluation of the two quantities. To ensure fairness, CPCG was initialized using 
the first evaluated point of GS. GS reaches to within $5\%$ and $0.5\%$ of the bound in 8 and 10 
iterations, and requires 110 iterations for convergence. In contrast, CPCG takes 5, 6 and 
64 iterations for $5\%$ and $0.5\%$ of the bound, and convergence respectively. The reason for the
slow convergence of GS is due to the oscillations occurring near the end, as this method does not
perform a methodical search across the constrained space, unlike CPCG. Clearly, CPCG is far superior 
to this method. As seen in Figure \ref{fig:trajectory}, both methods converge non-monotonically to 
the true CRB. 

\begin{table}[t]
\begin{center}
\begin{tabular}{|p{1.5cm}|p{1.1cm}|p{0.8cm}|p{0.8cm}|p{1.8cm}|}
\hline
\small \centering \textbf{Alg.} & \centering \small $\boldsymbol{\rho}$ &
\centering \small $\boldsymbol{5\%}$ & \centering\small $\boldsymbol{0.5\%}$ & \centering 
\small \textbf{Convergence} \tabularnewline
\hline \hline
%\centering CMM-R & \centering 0.9999 & \centering 16,524 & \centering 31,192 & \centering 93,475
%\tabularnewline 
\centering CMM-DD & \centering 0.9998 & \centering 12,935 & \centering 23,549 & \centering 69,026
\tabularnewline 
\centering CMM-CF & \centering 0.9986 & \centering 161 & \centering 407 & \centering 8,487
\tabularnewline 
%\centering GP & \centering 0.9988 & \centering  & \centering & \centering 
%\tabularnewline 
\centering CPCG & \centering 0.8715 & \centering  5 & \centering 7 & \centering 48 
\tabularnewline 
\hline
\end{tabular}
\end{center}
\caption{Asymptotic and finite convergence properties of the iterative algorithms}
\label{table:alg_performance}
\end{table}

\section{Conclusion}
\label{sec:conc}

In this paper, we revisit the problem of computing submatrices of the CRB. We show that computation
of these submatrices are related to a quadratic matrix program. Due to the properties of the 
FIM and the convexity of the quadratic matrix program, we can compute the
submatrices with efficient algorithms from convex optimization literature. We further show how the
framework here easily extends to the case when the FIM is singular, and when
parameter constraints are present. We then apply the algorithms on a constrained optimization 
problem, showing that the computation of these bounds can be evaluated efficiently for important
signal processing problems. Future work includes exploring more algorithms for evaluation that 
may possess faster convergence rates and testing on other constrained problems.

\appendix

\subsection{Derivation of the Optimal Solution of \eqref{eq:qmp}}
\label{app:opt_qmp}

The derivation relies on the relations $\nabla_{\bX} \tr(\bX^\T \bJ \bX) = 2 \bJ\bX$ and 
$\nabla_{\bX} \tr(\bB^\T\bX) = \bB$. Using these relations, the gradient of the problem is $\bJ \bX - 
\bB$. Setting this to $\mathbf{0}$, we obtain the optimal solution.

\subsection{Derivation of \eqref{eq:opt_constrained}}
\label{app:constrained}

Since $\bJ \in \Spsd$, the objective is a convex function, and the constraints are
linear, thus, the optimization problem remains convex. By using Lagrangian multipliers $\bZ$, the 
optimal solution obeys $\bJ \bX - \bB - \bH^\T \bZ = \mathbf{0}$ and $\bH^\T \bX = \mathbf{0}$. 
This implies that the feasible solutions of $\bX$ has the structure $\bX = \bU\bY$, where $\bY 
\in \Real^{(n-p)\times m}$, as solutions must lie in the cokernel of $\bH$ and the range space 
of $\bU$. Proceeding in this fashion, we obtain $\bJ\bU\bY = \bB$. Multiplying by $\bU^\T$ 
on both sides, we then get $\bY^\star = (\bU^\T \bJ\bU)^{-1}\bU^\T\bB$. Multiplying $\bY^\star$ 
again by $\bU$, we obtain the optimal solution $\bX^\star$. In the case of nonsingular $\bJ$, 
one can choose $\bU = \id{n} - \bJ^{-1} \bH (\bH^\T\bJ^{-1}\bH)^{+}\bH^\T$ as it
lies in the cokernel of $\bH$ and is orthogonal (see details in \cite{Gorman90CRB}). Then,
$\bX^\star$ is equivalent to \eqref{eq:opt_constrained_ns}. 

%\be
%\bJ\bU\bY = \bB.
%\label{eq:optimal_constrained}
%\ee

\subsection{Formulation of \eqref{eq:opt_sketch}}
\label{app:opt_sketch}

Consider the objective function once the inner minimization problem is solved. The objective 
function yields $-\tfrac{1}{2}\lbrack \I^{+}(p')\rbrack_{kk}$, for some particular rate sampling
rate $p'$. Now, $p^{\star}_k$ is the optimal value if and only if $-\tfrac{1}{2}\lbrack 
\I^{+}(p^{\star}_k)\rbrack_{kk} > -\tfrac{1}{2}\lbrack \I^{+}(p')\rbrack_{kk}$ for all other $p'
\ne p^{\star}_k$. By maximizing the objective function over $p$, we solve for $p^{\star}_k$.

\subsection{Derivation of the Constrained MM}

We prove the result for the vector case. Similar derivation applies for the matrix case. As discussed in Section 
\ref{sec:algo}, we use $Q(\bx;\bxi{k}) := \frac{1}{2}(\bx - \bxi{k})^\T (\bP - \bJ) (\bx - \bxi{k})$. Then,
the task is to minimize 
\ben
G(\bx;\bxi{k}) := \frac{1}{2}\bx^\T \bJ \bx - \bb^\T\bx + Q(\bx;\bxi{k}).
\een
subject to the constraint $\bH^\T\bx = \mathbf{0}$.

Using the method of Lagrangian multipliers \cite{Boyd04Opt}, we construct the Lagrangian, with multipliers $\bmu \in 
\Real^p$,
\be
L(\bx,\bmu;\bxi{k}) = G(\bx;\bxi{k}) + \bmu^\T \bH^\T \bx.
\label{eq:lagrangian}
\ee
At the optimal point, there are two equations to satisfy:
\begin{align*}
\nabla_{\bx} L(\bx,\bmu;\bxi{k}) &= \bP \bx + (\bP - \bJ)\bxi{k} - \bb + \bH\bmu = \mathbf{0}, \\
\nabla_{\bmu} L(\bx,\bmu;\bxi{k}) &= \bH^\T\bx = \mathbf{0}.
\end{align*}
Solving both equations, we have
\ben
\bmu^{(k+1)} = (\bH^\T \bP^{-1}\bH)^{-1} \bH^\T \Big( (\id{n} - \bP^{-1}\bJ) \bxi{k} - \bP^{-1}\bb \Big) ,
\een
which exists, since we choose $\bP \in \Spd$. Finally,
\begin{align*}
\bxi{k+1} &= (\id{n} - \bP^{-1} \bJ)\bxi{k} - \bP^{-1}\bb - \bH \bmu^{(k+1)}\\
&= \bT \Big( (\id{n} - \bP^{-1}\bJ) \bxi{k} - \bP^{-1}\bb \Big).
\end{align*}
where $\bT = \id{n} - \bH (\bH^\T \bP^{-1}\bH)^{-1} \bH^\T$ is a projection operator.
Note its similarity to the basic Jacobi iteration, except with the projection $\bT$ to account for the 
parameter constraints.

\small

%\nocite{Harville97MtxStat}

\bibliography{TuneBib,TunePubBib}
\bibliographystyle{abbrv}

%%%%%%%%%%%%%%%%%%%%%%%%%%%%%%%%%%%%%%%%%%%%%%%%%%%%%%%%%%%%%%%%%%%%%%%%%
\end{document}